\documentstyle[a4,12pt]{article}
\begin{document}
\pagestyle{empty}
\begin{titlepage}
\rightline{IC/94/327}
\rightline{UTS-DFT-94-15}
\rightline{October 1994}
\vspace{2.0truecm}
\begin{center}
\begin{large}
{\bf Search for new physics indirect effects in $e^+e^-\to W^+W^-$
at linear colliders with polarized beams}
\end{large}
\vspace{2cm}

{ A. A. Babich, A. A. Pankov\hskip 2pt\footnote{
E-mail: PANKOV@GPI.GOMEL.BY}
}\\[0.5cm]
{\it Gomel Polytechnical Institute, Gomel, 246746 Belarus, CIS}\\[1cm]
\vspace{3mm}

{  N. Paver\hskip 2pt\footnote{Also supported by the Italian
Ministry of University, Scientific Research and Technology (MURST).}
}\\[0.5cm]
{\it Dipartimento di Fisica Teorica, Universit\`{a} di Trieste, 34100
Trieste, Italy}
\\
{\it Istituto Nazionale di Fisica Nucleare, Sezione di Trieste, 34127 Trieste,
Italy}\\[1cm]
\end{center}

\vspace{1.5cm}

\begin{abstract}
\noindent
We discuss the potential of a $0.5\hskip 2pt TeV$ linear collider to
explore manifestations of extended (or alternative) electroweak models
of current interest, through measurements of the reaction
$e^+e^-\to W^+W^-$ with both initial and final states polarization.
Specifically, we consider the possibility to put stringent
constraints on lepton mixing (or extended lepton couplings) and $Z-Z^\prime$
mixing, showing in particular the usefulness of polarization in
order to disentangle these effects.
\end{abstract}
\end{titlepage}
\pagestyle{plain}
\noindent
It is generally believed that, although tremendously successful from the
phenomenological point of view, the Standard Model (SM) should not be
considered as the ultimate theory. Most proposed schemes attempting to
address the conceptual
problems of the SM predict the existence of additional building blocks,
namely new heavy gauge bosons and fermions. These extra degrees of freedom
might be too heavy for direct production and, in high energy reactions,
they could manifest themselves only as ``indirect'' effects
induced by mixings with the conventional SM fermions and bosons. In general,
such mixing effects reflect the underlying extended gauge symmetry and/or
the Higgs sector of the model.\par
Studies of high energy $e^+e^-$ annihilation can give an opportunity to make
precise tests of mixing effects. In $e^+e^-\to {\bar f}f$, one can
search for indirect effects at the standard $Z$ resonance \cite{langacker}-
\cite{altarelli}.
At the higher energies of planned linear colliders, this reaction should be
convenient mostly for the direct search of these new matter particles
\cite{desy1}-\cite{desy3}, if their production is allowed.\par
Conversely, at linear colliders, the process
\begin{equation}e^++e^-\to W^++W^-, \label{reaction}\end{equation}
should be quite sensitive to indirect new physics effects
\cite{pankov1,pankov2}, which can destroy the SM
gauge cancellation among the different contributions, and
hence cause deviations of the cross section from the SM prediction which can
increase with the $CM$ energy. In fact,
considering the cross sections for longitudinally polarized initial $e^-e^+$
beams, namely $\sigma^{LR}$ and $\sigma^{RL}$, significant improvements
of the present limits on lepton mixing and $Z$-$Z^\prime$ mixing could be
obtained. In this paper, we would like to extend the
analyses of lepton and $Z$-$Z^\prime$ mixing presented in
\cite{pankov1,pankov2}, and discuss the role of final $W^+W^-$
polarization measurements (combined with initial state polarization) in
order to further improve the bounds. In addition, we shall see that
combining initial and final states polarization provides a simple way to
derive separate bounds for lepton mixing and $Z$-$Z^\prime$ mixing. \par
Although the discussion could be done quite in general, we believe it more
interesting and useful to present this kind of analysis, and quantitatively
derive the bounds, making reference to some specific extended
(or alternative) electroweak models of current interest.\par
Specifically, for the leptons we shall limit ourselves to the cases listed
in Table 1, where the considered new fermions are either doublets or
singlets under the gauge symmetry $SU(2)$.
\begin{table}[b]
\centering
\begin{tabular}{|c|c|c|c|}
\hline
Leptons & Vector doublets & Vector singlets & Mirror fermions \\ \hline
$SU(2)$  & ${N^0\choose E^0}_L$ & $N^0_L,\ N^0_R$
& $N^0_L,\ E^0_L$ \\
structure  & ${N^0\choose E^0}_R$ & $E^0_L,\ E^0_R$
& ${N^0\choose E^0}_R$ \\ \hline
\end{tabular}
\caption{$SU(2)$ assignments for new leptons, $E$ and $N$ refer to
electric charge -$1$ and $0$ respectively. The superscript ``0'' means
weak eigenstates. }\label{tab:tab1}
\end{table}
We also assume that the new, ``exotic'' fermions only mix with the
standard ones within the same family (the electron and its neutrino being
the ones relevant to process (\ref{reaction})), which assures the absence of
tree-level generation changing neutral currents.
The needed fermion mixing formalism has been introduced, e.g.,
in \cite{pankov2}. Basically, denoting
by $\nu$, $e$, $N$ and $E$ the mass eigenstates, the neutral current couplings
of leptons to $Z$ and $Z^\prime$ can be written, respectively, as
\begin{equation}\ g_a^e=g_a^{e^0}c_{1a}^2+g_a^{E^0}s_{1a}^2;
\qquad\qquad\ \ \ \ \
g_a^{\prime e}=g_a^{\prime e^0}c_{1a}^2+g_a^{\prime E^0}s_{1a}^2,
\label{g}\end{equation}
where $f^0\equiv e^0,\hskip 2pt E^0$ are gauge-eigenstates, and
$a=L,\hskip 2pt R$. Moreover, in Eq. (\ref{g}) $c_{1a}=\cos\psi_{1a}$ and
$s_{1a}=\sin\psi_{1a}$, with $\psi_{1a}$ the mixing angle between the two
charged leptons.\par
The charged current couplings are given by:
\begin{eqnarray}G_L^{\nu}&=&c_{1L}c_{2L}-2 T_{3L}^Es_{1L}s_{2L};\qquad\ \ \
G_R^{\nu}=-2 T_{3R}^Es_{1R}s_{2R},\nonumber \\
G_L^{N}&=&-s_{2L}c_{1L}-2 T_{3L}^Ec_{2L}s_{1L};\qquad
G_R^{N}=-2 T_{3R}^Ec_{2R}s_{1R},\label{G}\end{eqnarray}
and, analogously to (\ref{g}), $c_{2a}=\cos\psi_{2a}$ and
$s_{2a}=\sin\psi_{2a}$ refer to the mixing between the neutral leptons.\par
In turn, $Z$-$Z^\prime$ mixing is introduced through the relation
\begin{equation}{Z_1\choose Z_2}=
\pmatrix{\cos\phi&\sin\phi\cr -\sin\phi&\cos\phi}{Z\choose Z^{\prime}},
\label{phi}\end{equation}
where $Z,\ Z^{\prime}$ are weak-eigenstates, $Z_1,\ Z_2$ are mass-eigenstates
and $\phi$ is the $Z$-$Z^{\prime}$ mixing angle.\par
As a class of models where lepton mixing and $Z$-$Z^\prime$ mixing can be
simultaneously present, we will consider in the sequel the case of $E_6$
models. In these extended schemes, the fermion couplings to $Z$ are
the familiar SM ones:
\begin{equation}
g_a^{f^{0}}=\left(T_{3a}^{f^{0}}-Q_{em,a}^{f^{0}}s_W^2\right)\hskip 2pt g_Z,
\label{gaf}\end{equation}
with $s_W^2=\sin^2\theta_W$ and $g_Z=1/s_Wc_W$, and the couplings to
$Z^\prime$ are:
\begin{eqnarray}g_L^{{\prime}e^{0}}&=&\left(3A+B\right)\hskip 1pt g_{Z^\prime}
\hskip 2pt; \qquad\ \ \ \ \
g_R^{{\prime}e^{0}}=\left(A-B\right)\hskip 1pt g_{Z^\prime}\nonumber \\
g_L^{{\prime}E^{0}}&=&-\left(2A+2B\right)\hskip 1pt g_{Z^\prime}\hskip 2pt;
\qquad g_R^{{\prime}E^{0}}=\left(-2A+2B\right)\hskip 1pt g_{Z^\prime}
\hskip 2pt.\label{AB}\end{eqnarray}
In Eq. (\ref{AB}): $g_{Z^\prime}=g_Z\hskip 1pt s_W$, $A=\cos\beta/2\sqrt 6,
\ B={\sqrt{10}}\sin\beta/12$, with $\beta$ specifying the orientation of the
$U(1)^{\prime}$ generator in the $E_6$ group space \cite{rizzo}. The most
commonly considered models are $Z^\prime_\chi$, $Z^\prime_\psi$ and
$-Z^\prime_\eta$ models, which are specified by $\beta=0$, $\pi/2$ and
$\left(\pi-arctan\sqrt{5/3}\right)$, respectively.\par
Finally, taking Eq. (\ref{phi}) into account, the leptons neutral current
couplings to $Z_1$ and $Z_2$ are, respectively:
\begin{equation} g_{1a}^{f}=g_a^{f}\cos\phi+g_a^{{\prime}f}\sin\phi
\hskip 2pt;
\qquad\qquad g_{2a}^{f}=-g_a^{f}\sin\phi+g_a^{{\prime}f}\cos\phi.
\label{gaffi} \end{equation}
Obviously, the SM case is reobtained when all (fermion and gauge boson)
mixing angles are put equal to zero.\par
The cross section of process (\ref{reaction}) can be expressed in general as
\begin{eqnarray}\frac{d\sigma\left(P_1,\hskip 2pt P_2\right)}{d\cos\theta}
&=&\frac{1}{4}\left[\left(1+P_1\right)\cdot \left(1-P_2\right)\hskip 2pt
\frac{d\sigma^{RL}}{d\cos\theta}+\left(1-P_1\right)\cdot \left(1+P_2\right)
\hskip 2pt\frac{d\sigma^{LR}}{d\cos\theta}\right.\nonumber \\
&+&\left.\left(1+P_1\right)\cdot\left(1+P_2\right)\hskip 2pt
\frac{d\sigma^{RR}}
{d\cos\theta}+\left(1-P_1\right)\cdot\left(1-P_2\right)\hskip 2pt
\frac{d\sigma^{LL}}{d\cos\theta}\right],\label{gcr}
\end{eqnarray}
where $P_1$ ($P_2$) are less than unity, and represent the actual degrees of
longitudinal polarization of $e^-$ ($e^+$).\par
The relevant polarized differential cross sections for
$e^-_ae^+_b\to W^-_{\alpha}W^+_\beta$ can be written as
\begin{equation}
\frac{d\sigma^{ab}_{\alpha\beta}}{d\cos\theta}=
C\cdot \sum_{k=0}^{k=2} F_k^{ab}\hskip 2pt{\cal O}_{k\,\alpha\beta}.
\label{A1}\end{equation}
Here $C=\pi\alpha^2_{e.m.}\beta_W/2s$, with $\beta_W=(1-4M_W^2/s)^{1/2}$
the $W$ velocity in the CM frame, and the helicities of the initial
$e^-e^+$ and final $W^-W^+$ states are labeled as
$ab=(RL,\hskip 2pt LR,\hskip 2pt LL,\hskip 2pt RR)$
and $\alpha\beta=(LL,\hskip 2pt TT,\hskip 2pt TL)$, respectively.
The ${\cal O}_k$ are functions of the kinematical variables which characterize
the various possibilities for the final $W^+W^-$ polarizations
($TT,\hskip 2pt LL,\hskip 2pt TL+LT$ or the sum over all $W^+W^-$ polarization
states for unpolarized W's). The $F_k$ are combinations of coupling constants
including the two kinds of mixings.\par
For the $LR$ case we have
\begin{eqnarray}
F_0^{LR} & = & \frac{1}{16s^4_W}\hskip 2pt\left[\left(G_L^{\nu}\right)^2+
r_N \left(G_L^N\right)^2\right]^2, \nonumber \\
F_1^{LR} & = & 2\left[1-g_{WWZ_1}g_{1L}^e\cdot\chi_1-
g_{WWZ_2}g_{2L}^e\cdot\chi_2\right]^2,\nonumber \\
F_2^{LR} & = & -\frac{1}{2s^2_W}\hskip 2pt \left[\left(G_L^{\nu}\right)^2+
r_N \left(G_L^N\right)^2\right]\hskip 2pt
\left[1-g_{WWZ_1}g_{1L}^e\cdot\chi_1-
g_{WWZ_2}g_{2L}^e\cdot\chi_2\right],\label{flr}\end{eqnarray}
where the $\chi_j$ are the $Z_1$ and $Z_2$ propagators, i.e.
$\chi_j=s/(s-M_j^2+iM_j\Gamma_j)$. Also, $r_N=t/(t-m_N^2)$, with
$t=M_W^2-s/2+s\cos\theta\hskip 1pt\beta_W/2$, and $m_N$ is the neutral
heavy lepton mass. The term proportional to $r_N$ represents the
neutral heavy lepton exchange diagram in the $t$-channel, which
accompanies the $\nu$-exchange. Finally, in Eq. (\ref{flr}),
$g_{WWZ_1}=\cot\theta_W\cos\phi$ and $g_{WWZ_2}=-\cot\theta_W\sin\phi$.\par
The $RL$ case is simply obtained from Eq. (\ref{flr}) by exchanging
$L\leftrightarrow R$. Also, for the $LL$ and $RR$ cases there is only
$N$-exchange:
\begin{eqnarray}
F_0^{LL} = F_0^{RR} = \frac{1}{16s^4_W}\hskip 2pt r_N^2 \hskip 2pt
\left(G_L^N G_R^N\right)^2.
\label{fll}\end{eqnarray}
Eq. (\ref{flr}) is obtained in the approximation where
the imaginary parts of the $Z_1$ and $Z_2$ boson propagators are neglected.
Accounting for this  effect requires the replacements $\chi_j\to Re\chi_j$ and
$\chi_j^2\to\vert\chi_j\vert^2$ ($j=1,2$) in the right-hand side of
Eq. (\ref{flr}).\par
Concerning the ${\cal O}_{k\,\alpha\beta}$ in Eq. (\ref{A1}), for the
cross section
${\displaystyle{\frac{d\sigma(e^-e^+\to W^-_LW^+_L)}{d\cos\theta}}}$
with any initial polarization, we have
(with $\vert\vec p\vert=\sqrt{s}\beta_W/2$):

\begin{eqnarray}
{\cal O}_{0,LL}&=&\frac{s\hskip 2pt \sin^2\theta}{4t^2M^4_W}\left[s^3(1+
{\cos^2\theta})-4M_W^4(3s+4M_W^2)-
4(s+2M_W^2)\vert\vec p\vert s\sqrt{s}\cos\theta\right], \nonumber \\
{\cal O}_{1,LL} & = & \frac{s^3-12sM_W^4-16M_W^6}{8sM_W^4}\sin^2\theta,
\nonumber \\
{\cal O}_{2,LL} & = & \frac{1-{\cos^2\theta}}{t}\left[\frac{\vert\vec p\vert s
\sqrt {s}(s+2M_W^2)}{2M_W^4}\cos\theta-\frac{s^3-12sM_W^4-16M_W^6}
{4M_W^4}\right].\label{oll}\end{eqnarray}
For the transverse ($TT$) cross section
${\displaystyle{\frac{d\sigma(e^+e^-\to W^+_TW^-_T)}{d\cos\theta}}}$ we have:
\begin{eqnarray}
{\cal O}_{0,TT} & = & \frac{4s}{t^2}\left[s(1+{\cos^2\theta})-2M_W^2-
2\vert\vec p\vert\sqrt{s}\cos\theta\right]\sin^2\theta, \nonumber \\
{\cal O}_{1,TT} & = & \frac{4{\vert\vec p\vert}^2}{s}\sin^2\theta,
\nonumber \\
{\cal O}_{2,TT} & = & \frac{\sin^2\theta}{t}\left[4\vert\vec p\vert
\sqrt{s}\cos\theta-8{\vert\vec p\vert}^2\right].\label{ott}\end{eqnarray}
For the production of one longitudinal plus one transverse
vector boson $(TL+LT)$ we have:
\begin{eqnarray}
{\cal O}_{0,TL} & = & \frac{2s}{t^2M^2_W}\left[s^2(1+\cos^4\theta)-
4\vert\vec p\vert\sqrt{s}\ cos\theta(4{\vert\vec p\vert}^2+
s\cos^2\theta)+\right. \nonumber \\
  & & \left. 4M_W^4(1+\cos^2\theta)+2s(s-6M_W^2)\cos^2\theta
-4sM_W^2 \right], \nonumber \\
{\cal O}_{1,TL} & = & \frac{4{\vert\vec p\vert}^2}{M_W^2}(1+{\cos^2\theta}),
\nonumber \\
{\cal O}_{2,TL} & = &
\frac{4\vert\vec p\vert\sqrt{s}}{tM_W^2}\left[(4{\vert\vec p\vert}^2+
s\cos^2\theta)\cos\theta-
2{\vert\vec p\vert}\sqrt{s}(1+\cos^2\theta)\right].\label{otl}\end{eqnarray}
Finally, in the case of equal helicities  of initial electron and
positron beams ($LL$ and $RR$):
\begin{eqnarray}
{\cal O}_{0,LL} & = & 4 \hskip 2pt\frac{s\hskip 2pt m_N^2}{M_W^4},\nonumber \\
{\cal O}_{0,TT} & = & 8 \hskip 2pt\frac{s\hskip 2pt m_N^2 }{t^2}\hskip 2pt
(1+\cos^2\theta),\nonumber \\
{\cal O}_{0,TL} & = & 2\hskip 2pt (1+\beta^2_W)\hskip 2pt
\left(\frac{m_N}{M_W}\hskip 2pt \frac{s}{t}\right)^2
\sin^2\theta.\label{llrr}\end{eqnarray}
\par In order to assess the sensitivity of the different cross sections to
ordinary lepton-exotic lepton mixing and to $Z$-$Z^\prime$ mixing, we
introduce the deviation from the SM cross section
$\Delta\sigma=\sigma-\sigma_{SM}$, where $\sigma\equiv\sigma(z_1,z_2)=
\int_{z_1}^{z_2}\left(d\sigma/dz\right)dz$ ($z=\cos\theta$), and
$z_1,\hskip 2pt z_2$ specify the kinematical range experimentally allowed.
Then, we define a $\chi^2$ function
\begin{equation}\chi^2=\left(\frac{\Delta\sigma}{\delta\sigma_{SM}}\right)^2,
\label{chisquare}\end{equation}
with $\delta\sigma_{SM}$ the accuracy experimentally obtainable on
$\sigma(z_1,z_2)_{SM}$. Including both statistical and systematical errors,
${\delta\sigma_{SM}=\sqrt{(\delta\sigma_{stat})^2+(\delta\sigma_{syst})^2}}$,
where $(\delta\sigma/\sigma)_{stat}=1/\sqrt{L_{int}\varepsilon_W\sigma_{SM}}$
with  $L_{int}$ the integrated luminosity and $\varepsilon_W$ the
efficiency for $W^+W^-$ reconstruction in the considered polarization state.
For that we take the channel of lepton pairs ($e\nu+\mu\nu$) plus two hadronic
jets, which corresponds to $\varepsilon_W\simeq 0.3$
\cite{frank}-\cite{anlauf}.\footnote{Actually, this reconstruction efficiency
might be somewhat smaller, depending on the detector \cite{frank}. On the
other hand, for our estimates we have taken a rather conservative choice for
the integrated luminosity, while
recent progress in machine design seems to indicate that quite larger
values are attainable \cite{settles} and can compensate for the reduction of
$\varepsilon_W$.} The criterion we shall follow to derive bounds on
the mixing angles will be to impose that
$\chi^2\leq\chi^2_{crit}$, where $\chi^2_{crit}$ is a number which specifies
a chosen confidence level and in principle can depend on the details of the
analysis. \par
For simplicity, we start our analysis by considering the case where there is
lepton mixing only, and no $Z$-$Z^{\prime}$ mixing. Present limits on
$s_1^2$ and $s_2^2$ are in general less than $10^{-2}$
\cite{nardi2}, so that we can expect that retaining only the terms
of order $s_{1a}^2$, $s_{2a}^2$ and $s_{1a}s_{2a}$ in $\Delta\sigma$ should
be an adequate approximation. Taking Eqs. (\ref{g}), (\ref{G}) and
(\ref{gaf}) into account, the lepton couplings needed in Eq. (\ref{gcr})
to this approximation, for the models of interest here, are collected in
Table \ref{tab:tab2}.\par
\begin{table} [hb]
\centering
\begin{tabular}{|c|c|c|}
\hline
Vector doublets & Mirror fermions & Vector singlets \\
\hline
$g_L^e=g_L^{e^0}$ &
$g_L^e=g_L^{e^0}+\frac{1}{2}s^2_{1L}$ &
$g_L^e=g_L^{e^0}+\frac{1}{2}s^2_{1L}$ \\
$g_R^e=g_R^{e^0}-\frac{1}{2}s^2_{1R}$ &
$g_R^e=g_R^{e^0}-\frac{1}{2}s^2_{1R}$ &
$g_R^e=g_R^{e^0}$ \\
$G_L^{\nu}=G_L^{\nu^0}-\frac{1}{2}\left(s_{1L}-s_{2L}\right)^2$ &
$G_L^{\nu}=G_L^{\nu^0}-\frac{1}{2}\left(s_{1L}^2+s_{2L}^2\right)$ &
$G_L^{\nu}=G_L^{\nu^0}-\frac{1}{2}\left(s_{1L}^2+s_{2L}^2\right)$ \\
$G_R^{\nu}=s_{1R}\hskip 2pt s_{2R}$ &
$G_R^{\nu}=s_{1R}\hskip 2pt s_{2R}$ &
$G_R^{\nu}=\hskip 2pt 0$ \\
$G_L^N=s_{1L}-s_{2L}$ &
$G_L^N=-s_{2L}$ &
$G_L^N=-s_{2L}$ \\
$G_R^N=s_{1R}$ &
$G_R^N=s_{1R}$ &
$G_R^N=0$ \\
\hline
\end{tabular}
\caption
{Lepton couplings for different lepton models in linear approximation on
$s_{1a}^2$, $s_{2a}^2$ and $s_{1a}s_{2a}$. Here $g_L^{e^0}=-\frac{1}{2}+s_W^2$,
$g_R^{e^0}=s_W^2$, $G_L^{\nu^0}=1$.}\label{tab:tab2}
\end{table}
The inputs in our analysis will be the following ones. We shall consider
as $RL$ or $LR$ the simplified situations $P_1=-P_2=P>0$ and $P_1=-P_2=-P$,
respectively, with $P=0.9$.\footnote{Such a value could be attainable
for electrons \cite{prescott}. In the case the positron beam was unpolarized,
Eq. (\ref{gcr}) shows that one would loose a factor $1/2$ in statistics, which
would affects by $\sqrt 2$ the bounds for the cases dominated by the
statistical uncertainty.} For the integrated luminosity we take
$L_{int}=20\hskip 2pt fb^{-1}$ as expected at the
planned $0.5\hskip 2pt TeV$ lineal colliders \cite{settles, treille}.
Concerning the experimentally accessible range for the scattering angle, we
assume a $10^\circ$ cut, i.e., $z_2=-z_1=0.98$. We will present the bounds
on the mixing angles at the two standard deviations level (or 95\% CL), which
for our analysis corresponds to $\chi^2_{crit}=4$. Also, we take
$(\delta\sigma /\sigma)_{syst}=2\%$, as currently assumed \cite{frank}.\par
As indicated in Table \ref{tab:tab2}, for the case of lepton vector doublets
$\delta\sigma^{RL}$ and $\delta\sigma^{LR}$ separately constrain $s_{1R}^2$
and ${\left(s_{1L}-s_{2L}\right)^2}$, respectively. On the other hand,
Table \ref{tab:tab2} and Eq. (\ref{flr}) show that that it is not possible
to constrain $s_{2R}^2$ in the adopted approximation for $\Delta\sigma$.
Analogously, one expects a drastically reduced sensitivity to $s_{1L}^2$
for mirror and vector singlet models, basically due to the cancellation
between the $g^e_L$ and the $G^{\nu}_L$ entries in Table \ref{tab:tab2}.\par
The bounds on $s_{1R}^2$ and ${\left(s_{1L}-s_{2L}\right)^2}$ for vector
doublet leptons are represented in Fig. 1, as functions of
the heavy neutral lepton mass $m_N$. Since we are concentrating on indirect
effects, for $m_N$ we have assumed a minimum value of $250\hskip 2ptGeV$.
Concerning the final $W^+W^-$ polarizations, it turns out that the most
restrictive limits (which can compete with current ones) result from the
$LL$ case. In Fig. 2 we report, for $m_N=300\hskip 2pt GeV$, the bounds on
$s_{1R}^2$ and $\left(s_{1L}-s_{2L}\right)^2$ from Fig. 1 along with the
area allowed by the cross sections for $RR$ and $LL$ initial polarizations
with, respectively, $P_1=P_2=\pm 0.9$ (and longitudinal $W$'s). As a
curiosity, if
a ``signal'' (namely, a deviation from the SM) was observed for $RR+LL$
initial polarizations, separate determinations of
$s_{1R}^2$, ${\left(s_{1L}-s_{2L}\right)^2}$ and $m_N$ would be possible from
the combination of the three cases $RL$, $LR$ and $RR+LL$.
Conversely, if no signal at all was observed, still consideration of the cross
section for initial $RR+LL$ might be useful to improve the limits allowed by
$RL$ and $LR$, for not too heavy $m_N$ ($m_N<130\hskip 2pt GeV$ for the
inputs assumed here),
because in this case the $RR+LL$ curve would be able to cross the region
allowed by the other two cross sections. For larger $m_N$ the limits on
$s_{1R}^2$ and ${\left(s_{1L}-s_{2L}\right)^2}$ are determined by solely the
$RL$ and $LR$ cross sections. In Fig. 1 we show also the bounds for the
limiting case $m_N\to\infty$. Indeed, in general the most stringent
constraints are obtained for larger values of $m_N$. This feature reflects
the decreasing behaviour of the heavy neutrino exchange contribution to
$\Delta\sigma$, which is proportional to $r_N$ and opposite in sign to the
other mixing effects (see Eq. (\ref{flr}) and Table 2). This leads
to larger $\Delta\sigma$ in Eq. (\ref{chisquare}) and correspondingly to
a better sensitivity on the mixing angles. In any case, the limits
cannot go below those corresponding to $m_N\to\infty$ if the luminosity is
fixed at the chosen value.\par
Turning to the other exotic lepton models, from Table \ref{tab:tab2} one
can notice the following relations for the deviations from the SM:
\begin{eqnarray}
\Delta\sigma^{LR}(\hbox{mirror}) & = & \Delta\sigma^{LR}(\hbox{singlet}),
\nonumber \\
\Delta\sigma^{RL}(\hbox{doublet}) & = & \Delta\sigma^{RL}(\hbox{mirror}),
\nonumber\\
\Delta\sigma^{RL}(\hbox{singlet}) & = & 0.\label{relat}\end{eqnarray}
Due to Eq. (\ref{relat}), in the case of no observed signal only three
kinds of bounds can be derived, instead of the six ones which could be
expected a priori. Specifically, in addition to the two ones for the vector
doublet model exhibited in Figs. 1 and 2, the third one results from the
initial $LR$ polarization and is relevant either to the mirror and the
singlet leptons. This is represented in Fig. 3 which, shows the corresponding
upper bound on $s_{2L}^2$ for these models.\par
We now discuss the case where there is $Z$-$Z^\prime$ mixing only, and no
lepton mixing. In this regard, we recall from Eq. (\ref{phi}) that
\begin{equation}\tan^2\phi=\frac{M_Z^2-M_1^2}{M_2^2-M_Z^2}, \label{tan}
\end{equation}
with $M_Z^2=M_W^2/\cos^2\theta_W$, and, since we are interested in indirect
effects, $M_1\ll E_{CM}\ll M_2$. In this case the deviations from the SM
arise from the fermion coupling constants in Eq. (\ref{gaffi}) and from
the $Z_1$ exchange amplitude which involves the $Z$ mass shift
$\Delta M=M_Z-M_1$ induced by the $Z$-$Z^\prime$ mixing angle $\phi$
\cite{pankov1}. Current limits on $\phi$ and $\Delta M$ from LEP are,
in general, $\vert\phi\vert< 0.01$ \cite{langacker}-\cite{altarelli} and
$\Delta M< 90\hskip 2pt MeV$ \cite{altarelli}.\par
In Fig. 4 we report the limits on $\phi$ as a function of the $E_6$ model
parameter $\cos\beta$, resulting from the cross section for both longitudinal
final $W$ (which turns out to be the most sensitive one to such mixing
effects), with the different initial beams polarizations. This figure shows
the complementary roles of the $RL$ and $LR$ cross sections
in limiting $Z$-$Z^\prime$ mixing over the full range of $\beta$. The
typical limits on $\vert\phi\vert$ are of the order of $1.5\times 10^{-3}$ to
$2.5\times 10^{-3}$, and for selected values of $\beta$ can considerably
improve the present situation \cite{langacker}-\cite{altarelli}.
The orders of magnitude in Fig. 4 could be directly guessed by
considering that the departure from the SM cross section relevant, e.g., to
$e^+_Le^-_R\to W^+_LW^-_L$ entering in Eq.(\ref{chisquare}) is:
$\Delta\sigma^{RL}_{LL}\propto\Delta F_1^{RL}$. In turn, $\Delta F_1^{RL}$
can be written, up to linear terms in $\phi$ and $\Delta M$, as
\begin{equation}
\Delta F_1^{RL}=-4\left(1-
\cot\theta_W \hskip 2pt g^e_R \hskip 2pt\chi_Z\hskip 1pt \right)
\hskip 2pt \cot\theta_W \hskip 2pt g^e_R \hskip 2pt\chi_Z\hskip 1pt
\left(\varepsilon_{mix}+\varepsilon_{int}+\varepsilon_{\Delta M}\right),
\label{a2}\end{equation}
where $\varepsilon_{\Delta M}=-2M_Z\Delta M/(s-M_Z^2)=-7.5\times 10^{-4}\hskip
2pt (\Delta M/GeV)$,
$\varepsilon_{mix}=\phi(g^{\prime e}_R/g^e_R)$ and
$\varepsilon_{int}=-\varepsilon_{mix}(\chi_2/\chi_Z)$.
The condition $\chi^2\leq\chi^2_{crit}=4$ gives the bound
\begin{equation}\vert\phi\vert<\frac{1}{2}\sqrt{\chi^2_{crit}}\hskip 4pt
\left(\frac{\delta\sigma_{SM}}{\sigma_{SM}}\right)
\left|{\frac{1-\cot\theta_W \hskip 2pt g^e_R \hskip 2pt\chi_Z\hskip 1pt}
{\cot\theta_W \hskip 2pt g^{\prime e}_R \hskip 2pt\chi_Z\hskip 1pt}}\right|.
\label{bound}\end{equation}
Eq. (\ref{bound}) gives, for example in the case of the $\chi$-model,
$\vert\phi\vert<3\times 10^{-3}$ with weak dependence
on $\Delta M\sim 90\hskip 2pt MeV$. Analogously, for $LR$ case one can simply
find $\vert\phi\vert<1.5\times 10^{-3}$.\par
Finally, we discuss the case where both lepton mixing
and $Z$-$Z^\prime$ mixing
occur, so that the leptonic coupling constants are as in Eq. (\ref{gaffi})
and the $Z_1,\hskip 2pt Z_2$ couplings to $W$ are as in Eq. (\ref{flr}).
In this case we can look for the regions allowed to the combinations
$(s_{1R}^2,\hskip 2pt\phi)$ and $((s_{1L}-s_{2L})^2,\hskip 2pt\phi)$, for
fixed $m_N$. Figs. 5 and 6 show, as typical examples, the results of this
analysis for the $\psi$-model and the $\chi$-model, respectively, with fixed
$m_N=300\hskip 2pt GeV$. As one can see, the shapes of the allowed regions
for the mixing angles are quite different for these two cases. From the
explicit calculation it turns out that this is due to the
different relative signs (depending on the value of $\beta$) between
lepton and $Z$-$Z^\prime$ mixing contributions to the deviations
$\Delta\sigma$. Specifically, for the $\psi$-model the coefficient of the
$\phi$-term has same or opposite sign with respect to the
lepton mixing-term (which is always negative) for the $RL$ and $LR$ initial
polarizations, respectively. On the contrary, for the $\chi$-model the above
signs are opposite to the lepton mixing-term for both $LR$ and $RL$ initial
polarizations.\par
Concerning Fig. 5 and the corresponding analysis for the $\psi$-model,
we should notice that the limits on the $Z$-$Z^\prime$ angle $\phi$ are
numerically quite consistent with those previously found by assuming
zero lepton mixing $s_{1R}^2=(s_{1L}-s_{2L})^2=0$, and numerically improve
the present findings. In turn, the bounds on the lepton mixing angles
$s_{1R}^2$ and $(s_{1L}-s_{2L})^2$ are
somewhat looser than in the case $\phi=0$ discussed above
(roughly, by a factor of two), but still numerically competitive with the
current situation. Finally,
we can remark that the cross sections
for longitudinal $W^+W^-$ production provide by themselves the most stringent
constraints for this model.\par
Different from the $\psi$-model, Fig. 6 shows that the cross sections for
longitudinal $W$-pair production must be supplemented by the measurement
of the $TL$ final state in order to obtain the best limits for the
$\chi$-model.
Also, for such model, in the range of negative $\phi$ the limit on
$Z$-$Z^\prime$ mixing is consistent with the one obtained in the case of
zero lepton mixing,  while the bound somewhat worsens for positive $\phi$.
However, these limits can be still considered as an
improvement over the present bounds. On the other hand, the constraints
on the lepton mixing angles are looser than those found in the case of
zero $Z$-$Z^\prime$ mixing. In this regard one should notice that there is
a strong correlation between lepton mixing and $Z$-$Z^\prime$ mixing
which could be very useful in order to test the $\chi$-model. Also, it turns
out that the shapes of the $RL$ and $LR$ allowed regions in Fig. 6 change
with the CM energy, giving different intersections among the relevant curves.
Thus, the combination of measurements at different energies should
considerably help to reduce of allowed regions. \par
The examples discussed above show the advantage of combining observables
corresponding to the various initial and final polarizations in order
to test physical effects beyond the Standard Model which occur through
deviations from the SM cross sections in process (\ref{reaction}). As
a small final remark,
we would mention that the procedure presented above could be
applied more extensively. A case where a similar analysis might prove
useful could be the strongly interacting electroweak
symmetry breaking model BESS \cite{casalbuoni1}-\cite{casalbuoni3}.
The non-standard parameters of this model, in its simplest version, are the
mass $M_V$ of a new heavy neutral boson, its gauge coupling
$g^{\prime\prime}$, and its direct coupling to fermions $b$.
As an example Fig. 7 shows the bounds, which might be obtained from the
combination of $LR$ and $RL$ cross sections for longitudinal $W$,
on ratio of gauge couplings $g/g^{\prime\prime}$ and $b$. Also in this case,
initial state polarization is found to help in reducing the region allowed to
the model parameters.
\newpage
\begin{center}{\bf Acknowledgements}\end{center}
\noindent
AAB and AAP acknowledge the support and the hospitality
of INFN-Sezione di Trieste and the International Centre for
Theoretical Physics, Trieste.
The work of (NP) has been supported in part by the
Human Capital and Mobility Programme, EEC Contract ERBCHRXCT930132.

\newpage
\section*{Figure captions}
\begin{description}

\item{\bf Fig.1} Upper bounds (95\% C.L.) on $s_{1R}^2$ and
$(s_{1L}-s_{2L})^2$ for vector doublet leptons vs. $m_N$,
from $e^-e^+\to W^-_LW^+_L$ and initial $LR$ and $RL$ polarization.
Here: $E_{CM}=0.5\hskip 3pt TeV$,
$L_{int}=20\hskip 2pt fb^{-1}$, $P_1=-P_2=0.9\ (RL)$, $P_1=-P_2=-0.9\ (LR)$.
The horizontal dashed (dotted) line indicate the bounds for the $LR$ ($RL$)
case with $m_N\to\infty$.

\item{\bf Fig.2} Upper bounds (95\% C.L.) on $s_{1R}^2$ and
$(s_{1L}-s_{2L})^2$ from $e^-e^+\to W^-_LW^+_L$ with polarized $RL$,
$LR$ and $LL+RR$ ($P_1=P_2=\pm 0.9$) initial beams, respectively,
and at $m_N=300\hskip 3pt GeV$. All other inputs are the same as in Fig. 1.

\item{\bf Fig.3} Upper bounds on $s_{2L}^2$ vs. $m_N$ for mirror (and
singlet) leptons from $e^-e^+\to W^-_LW^+_L$ with initial $LR$ polarization.
Same inputs as for Fig. 1. The horizontal dashed line indicates the bound
for the $LR$ case with $m_N\to\infty$.

\item{\bf Fig.4} Upper limits (95\% C.L.) for $\phi$ vs. the
$E_6$ model parameter $\cos\beta$ from longitudinal $W$ pair production.
Same inputs as for Fig. 1.

\item{\bf Fig.5} Allowed regions for the combinations
($s_{1R}^2,\hskip 2pt\phi$) and ($(s_{1L}-s_{2L})^2,\hskip 2pt\phi$)
for the $\psi$-model,  from $e^-e^+\to W^-_LW^+_L$ with $RL$ and $LR$ initial
polarization, $m_N=300\hskip 3pt GeV$. All other inputs as in Fig. 1.

\item{\bf Fig.6} Same as in Fig. 5, for the $\chi$-model,
from $e^-_Le^+_R\to W^-_LW^+_L$, area enclosed between solid lines;
from $e^-_Re^+_L\to W^-_LW^+_L$, area enclosed between dashed lines;
from $e^-_Le^+_R\to W^-_TW^+_L$, area enclosed between dotted lines.

\item{\bf Fig.7} Allowed region (95\% CL) for
($b,\hskip 2pt g/g^{\prime\prime}$) with $M_V=1\hskip 2pt TeV$,
from $e^-e^+\to W^-_LW^+_L$ with $RL$, $LR$ and unpolarized initial
beams.
\end{description}
\end{document}